\documentclass[preprint]{aastex}

\shorttitle{Fe Abundance in HH Shocks}
\shortauthors{B\"ohm \& Matt}

\slugcomment{Accepted by \pasp.}

\begin{document}

\title{An Approximate Determination of the Gas-Phase Metal Abundance
in Herbig-Haro Outflows and their Shocks}

\author{Karl-Heinz B\"ohm and Sean Matt}
\affil{Astronomy Department, University of Washington, Seattle WA 98195\\
bohm@astro.washington.edu, matt@astro.washington.edu}

\begin{abstract}

It is important to determine whether the observed bow shocks in the
working surfaces of Herbig-Haro outflows has lead to a destruction of
dust grains and consequently to a change in the gas phase metal
abundances (say of Fe) in the cooling regions of HH bow shocks.
Detailed studies are currently available for only 5 HH outflows
(Beck-Winchatz et al.\ 1996). This small number is due to the large
observational and theoretical effort required to determine metal
abundances in HH objects.

Information about metal abundances in more HH objects is badly needed.
We therefore use a very approximate method. We introduce a
``characteristic number,'' $A_{\rm me}$, whose definition is based
only on the often observed line fluxes of [Fe II] 5159, [Fe II] 7155,
[Ca II] 7291, as well as H$\alpha$ and H$\beta$.  These fluxes can
easily be determined from existing observations.  We find a good
correlation between $A_{\rm me}$ and the Fe abundance for the 5
well-studied HH objects. We use this correlation to determine
approximate values of the gas phase Fe abundance in 13 additional high
excitation and in 3 additional low excitation HH objects.

The results are the following: Of the 16 high excitation HH objects
studied, there are 6 which approximately agree with normal population
I abundance (i.e., no depletion due to dust formation).  The remaining
10 show some (very modest) Fe gas phase depletion which, however,
never gets larger than by a factor of 2.5.  This result is in
agreement with our qualitative expectations that fast shocks
efficiently destroy dust grains.  Of the 5 low excitation HH objects
studied, there are 4 which show a normal population I abundance
(strictly speaking, an even slightly higher abundance than this). This
is completely unexpected. In low excitation objects, one might expect
strong gas phase Fe depletion (showing the unchanged molecular cloud
composition), unless the matter has previously gone through shocks of
much higher shock velocities. We discuss this possible explanation and
the question of whether low excitation HH objects have a different
``history'' than usually assumed.

\end{abstract}


\section{Introduction}

It has been generally accepted that the Herbig-Haro outflows (jets)
from young stellar objects become optically visible through the shocks
(bow shocks and jet shocks, Mach disks) at their working surfaces
(Mundt \& Fried 1983; Hartmann \& Raymond 1984; B\"ohm \& Solf 1985;
Schwartz 1985; Mundt 1985; Hartigan, Raymond \& Hartmann 1987;
Reipurth 1989; Hartigan 1989; Solf \& B\"ohm 1991; Reipurth \&
Heathcote 1993; Eisloeffel, Mundt \& B\"ohm 1994; Raga 1995).  In
other words, in HH objects we see matter (if it is observed in a bow
shock) that originally comes from the molecular cloud and has recently
gone through the shock and through an essential part of the cooling
region of the shock.  Recently, it has also become clear that, in a
number of Herbig-Haro (HH) objects, we can identify ``internal working
surfaces'' and their bow shocks (Raga 1995; Reipurth \& Heathcote
1997; Reipurth et al.\ 1997; Raga \& Noriega-Crespo 1998).  That is,
in a number of HH objects (e.g., in HH 111 or in HH 34), the cloud
matter has passed through multiple shocks.

In molecular clouds (including the boundary regions of the clouds,
which are accessible to optical HH observations), we expect that the
gas phase abundances of elements like Fe or Ca are considerably
depleted (by a factor of on the order of 100) because of the formation
of dust grains (van Dieshoeck et al.\ 1993). If one takes into account
the fact that refractory elements like Fe and Ca can return to the gas
phase by the destruction of the dust grains in sufficiently strong
shocks ($v = 150$ km s$^{-1}$; e.g., McKee et al.\ 1987; Seab 1987;
Draine 1995), we might expect that HH objects will show different Fe/H
or Ca/H ratios that depend on the velocity of the shock through which
the visible matter has passed before it is observed.

The (admittedly very naive) expectation would be that ``low excitation
objects'' (slow shocks, $v \la 50$ km s$^{-1}$), would have rather
small gas phase Fe and Ca abundances because these elements remain in
the dust grains.  In ``high excitation objects'' (fast shocks, $v >
85$ km s$^{-1}$), we might expect the dust grains to be destroyed, at
least partially, by the shock.  The gas phase abundance of elements
like Fe or Ca would be high (solar?).  This hypothesis inspired
Beck-Winchatz, B\"ohm, \& Noriega-Crespo (1994, hereafter BBN94; 1996,
hereafter BBN96) to determine the Fe gas phase abundance for a few
objects.

In both papers the authors decided that it would be most convenient
(for the study of dust grain survival) to determine the Fe abundance
(they derived the abundance ratio Fe/S and Fe/O).  The effort required
for carrying out this program for a single HH object is sizable. It
requires reliable line spectrophotometry with fairly high spectral
resolution in a fairly large wavelength interval and the detailed
study of the statistical equlibrium of Fe (and a comparison element,
either O or S).  For this reason the program was carried out for only
5 HH objects, namely for HH 1, HH 7, HH 11, HH 43A, and HH 255
(Burnham's Nebula; BBN94, BBN96).

Many of their results are surprising and do not (in general) agree
with our above mentioned expectations.  First, the Fe abundance for
both HH1 and HH7 are essentially solar (with only 18\% and 3\%
deviation, respectively).  This is surprising because HH 1 is the
standard high excitation object, while HH 7 is one of the lowest
excitation objects known (so we expect a low Fe abundance).  Second,
the Fe abundance of the low excitation object HH 11 is (for unknown
reasons) somewhat higher (by a factor of 1.68) than the solar
abundance.  Finally, the authors detected some depletion of Fe in HH
43A (Fe/O about 38\% and Fe/S about 48\% of the solar value) and HH
255 (Fe/O about 71\% and Fe/S about 36\% of the solar value), though
these are high excitation objects. In judging these results we have to
be aware that HH 255 is a somewhat complicated HH object (see B\"ohm
\& Solf 1997; Solf \& B\"ohm 1999), to which some of the standard
assumptions (e.g., that the line emission is exclusively caused by a
bow shock in the working surface of the outflow) may be only partly
applicable.  Even if we exclude for the moment HH 255 from our
consideration, it is clear that the earlier results do not agree with
the simple expectations expressed above.  One way to explain why the
low excitation objects (HH 7 and HH 11) show no Fe depletion is to
assume that the shocked gas we observe has passed through a much
faster shock at some earlier time.
     
In order to clarify this problem, it is important to determine the gas
phase Fe abundance (and correspondingly the destruction of dust
grains) in a larger number of HH objects.  However, as discussed
above, this is very difficult to do in detail.  We therefore consider
it presently justified to try to determine preliminary, (very)
approximate values for the Fe abundance in a larger number of HH
objects, even if they are considerably less reliable than the earlier
determinations of BBN96.

First, we want to mention related studies by other authors.  Hartigan
et al.\ (1999, 2000) derived gas phase Fe abundance in HH objects
which approximately agree with BBN94 and BBN96. In their 1999 paper,
they studied the Fe II and [Fe II] spectrum for HH 47A in unusual
detail, covering the bow shock and the Mach disk separately.  Their
conclusion was that the gas phase Fe abundance is approximately solar.

It is also worth noting that there has been another recent study of
the gas phase Fe abundance which gave a somewhat lower abundance (on
the average by a factor of 3) for HH objects by Mouri \& Taniguchi
(2000).  They use a plane shock wave model to calculate the
theoretical line ratio [Fe II] 8617/[O I] 6300 in considerable detail
and compare this to observations of HH objects (which are not
individually listed).  We agree (of course) that HH objects are caused
by shock waves, but an attempt to explain in quantitative detail
individual spectra by a (plane or bow shock) model leads to
difficulties \citep{ragaea96,bg97}.  For this reason, the approach of
BBN94 and BBN96 was completely different.

BBN94 and BBN96 studied the HH objects individually and tried to
introduce (where possible) observationally determined parameters
(empirically determined densities, sizes, and electron temperatures of
the line emitting regions in HH objects).  They used different methods
to get information about the ionization.  In one method, for example,
BBN96 determined the abundance ratio 
\begin{equation}
	{{\rm N(Fe)}\over{\rm N(O)}} = 
	{{\rm N[Fe\: II] + N[Fe\: III]}\over
	{\rm N[O\: I] + N[O\: II] + N[O\: III]}} \nonumber
\end{equation}
(where N[X] is the number of [X] ions).  (This does not require
detailed theoretical information about the ionization.)  They also
based the derived abundance information on many lines (e.g., 42 [Fe
II] lines in HH 1).

Since the approaches of \citet{mt00} and BBN94/96 are completely
different (one based on a detailed, but basically very simple model,
the other on determining the empirical parameters for individual HH
objects), we presently cannot say which is more correct.  This will
depend on obtaining more insight into the detailed structure of
individual HH objects.  However, our following approach has been to
employ knowledge of individual objects, and consequently, we follow
the work of BBN96.

We proceed as follows: In \S \ref{method}, we introduce a
characteristic number which uses the line fluxes of only a very few
lines which are observed in reasonably many HH spectra (which are not
taken with the special purpose of determining the Fe abundance, see
Raga, B\"ohm, \& Canto 1996 for a reference to appropriate spectra).
We show that the characteristic number is well correlated with the
actual Fe abundance for the 5 HH objects for which the Fe abundance
has been determined in detail (BBN96).  In \S \ref{application}, we
use this correlation to determine approximately the gas phase Fe
abundance in the other HH objects.  Finally, in \S \ref{conclusions},
we discuss the enigmatic results and suggest possible explanations.

\section{An Approximate Method for Determining Gas-phase Fe Abundance 
\label{method}}

We want to develop a diagnostic method that uses only a few frequently
observed lines.  We will use the Fe lines [Fe II] 5159 and [Fe II]
7155, which have been observed in many objects. It would be nice to
add a third Fe line to the criterion. Since there are no other Fe
lines as frequently observed in the optical spectrum as these, we have
also used the [Ca II] 7291 line. This may appear somewhat
inconsistent, since we wish to determine the Fe abundance.  However,
one has to remember that the gas phase abundance of Ca behaves similar
to that of Fe (with regard to the dust grain destruction). (One also
has to keep in mind that we are constructing a characteristic
number which will be approximately correlated with the Fe abundance
but is not the Fe abundance itself.)  Using these lines, we define
\begin{equation} 
\label{ame}
A_{\rm me} \propto
	{{\rm F}_{\rm [Fe\: II]\; 5159}\over{\rm F}_{\rm H\beta}} +
	{{\rm F}_{\rm [Fe\: II]\; 7155}\over{\rm F}_{\rm H\alpha}}  +
	{{\rm F}_{\rm [Ca\: II]\; 7291}\over{\rm F}_{\rm H\alpha}} 
\end{equation}
where F$_{\rm [Fe II]\; 5159}$ is the measured flux of the [Fe II]
5159 line, and similarly for the [Fe II] 7155 and [Ca II] 7291 lines.
The [Fe II] 5159 line is measured relative to the H$\beta$ flux, while
[Fe II] 7155 and [Ca II] 7291 are relative to H$\alpha$. This is done
to keep the error which is introduced by the different reddening of
the HH objects small.  In order to get a feeling for the possible
error of this assumption, we use the reddening for HH 1
\citep{solfea88}.  In this case, E$_{\rm B - V} \sim 0.43$ (A$_{\rm V}
\sim 1.35$).  This leads to an error in F$_{\rm [Fe\: II]\; 5159}/{\rm
F}_{H\beta}$ of 15.6\% if we use the directly measured ratio instead
of the reddening corrected ratio.  For the case of F$_{\rm [Ca\: II]\;
7291}/{\rm F}_{\rm H\alpha}$, the corresponding error is 13.4\%.  We
choose a proportionality constant so that all $A_{\rm me}$ values are
given relative to the value for HH 1.  It would also be possible to
use the near IR [Fe II] lines (see, e.g., Gredel 1994), but the
influence of the reddening on the ratio of, e.g., [Fe II] 12570, 16439
to H$\alpha$ would cause additional problems.

\begin{figure}[t]
\plotone{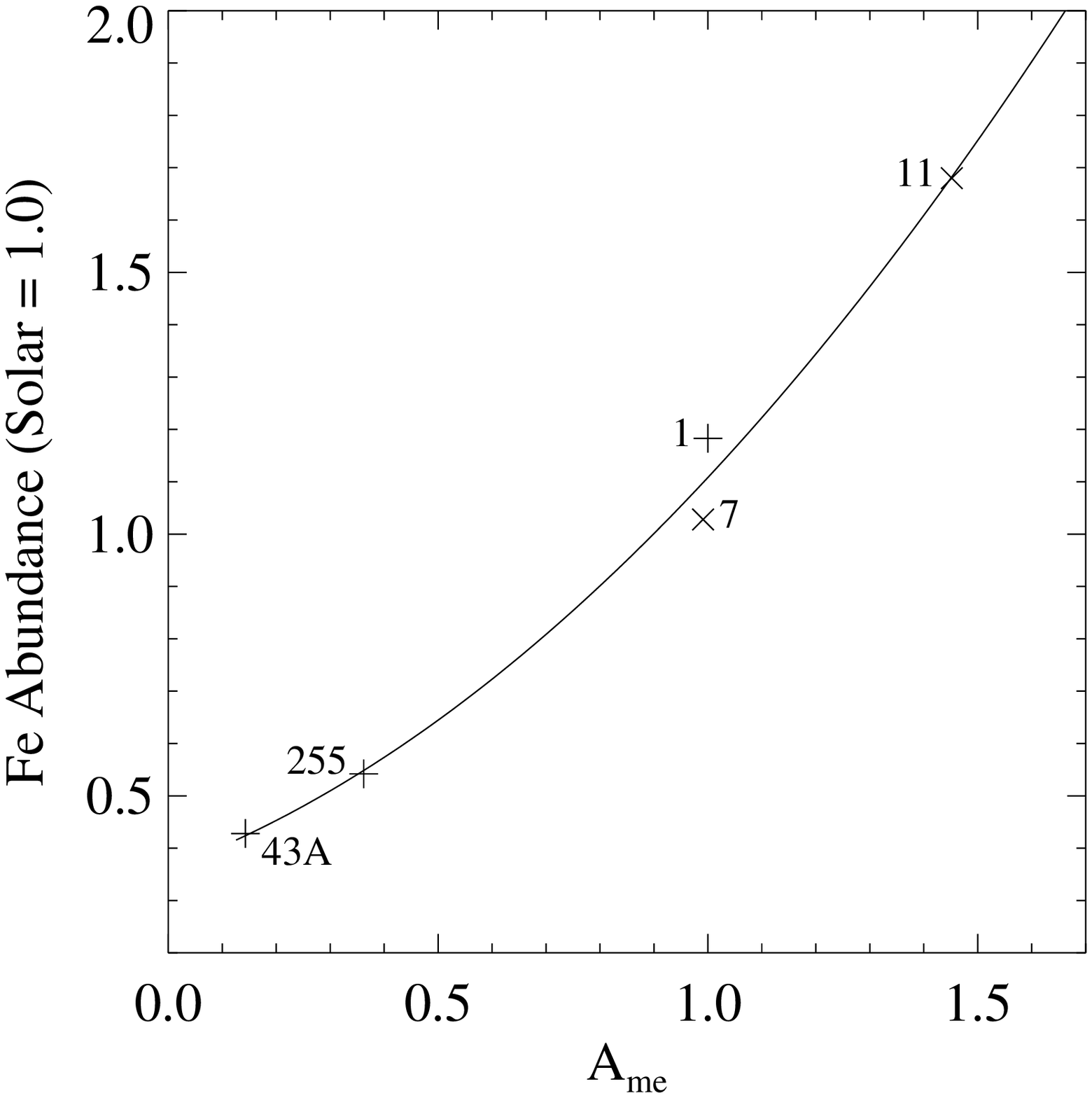}

\caption{Five HH objects (whose Fe abundance was determined in detail
by BBN96) indicate a possible correlation between the Fe abundance and
the characteristic number, $A_{\rm me}$ (see text).  High excitation
objects are designated by ``plus'' symbols, while ``x'' symbols
correspond to low excitation objects.  The HH designation is indicated
next to each symbol (the number and letter designation identifies the
HH object by the name assigned to it in Reipurth's 1999 catalogue).
The solid line is the least squares fit of a second order polynomial
to the data.  \label{fig1}}

\end{figure}

The question of whether this (somewhat arbitrary) definition of
$A_{\rm me}$ can be justified is best addressed by looking at the
correlation of $A_{\rm me}$ with the Fe gas phase abundances of the 5
HH objects whose abundances have been studied in detail (BBN94,
BBN96). In Figure \ref{fig1}, this correlation between $A_{\rm me}$
and the determined gas phase abundance of Fe is evident.  Here, the
iron abundance is normalized to 1 for the solar Fe abundance.  (This
corresponds to the abundance ratio N(Fe)/N(S) = 2.95 or N(Fe)/N(O) =
0.55, see Grevesse \& Anders 1991.)  Is it surprising that there is a
reasonable correlation between such a simple interpolation formula as
given in Equation \ref{ame} and the carefully determined Fe abundance?
One might have expected that the characteristics of an individual HH
object in question might have a sizable influence on the relation
between $A_{\rm me}$ and the Fe abundance.  However, one has to keep
in mind that the high excitation objects all have relatively similar
shock structure (shock velocities between 85 and 115 km s$^{-1}$, see
B\"ohm \& Goodson 1997).  It is therefore apparent that, for these
objects, the change in $A_{\rm me}$ is mostly caused by a change in
gas phase Fe abundance, while changes of the model are much less
important.

It seems more surprising that there appears to be a continuous
extension of the (approximate) correlation from the high excitation
objects to the low excitation objects (which would not be the case if
the correlation depended only on the shock velocity).  However, in
this respect, it is important that (for reasons which are not fully
understood) there is a typical observed difference between low and
high excitation objects.  High excitation objects typically have a
pre-shock density of $\sim 10^2$ cm$^{-3}$, while low excitation
objects have a typical pre-shock density of $\sim 10^3$ cm$^{-3}$
\citep{bg97,ragaea96}.  This change may contribute to an almost
continuous extension of the relation between $A_{\rm me}$ and Fe
abundance (Fig. \ref{fig1}).

\begin{figure}[t]
\plotone{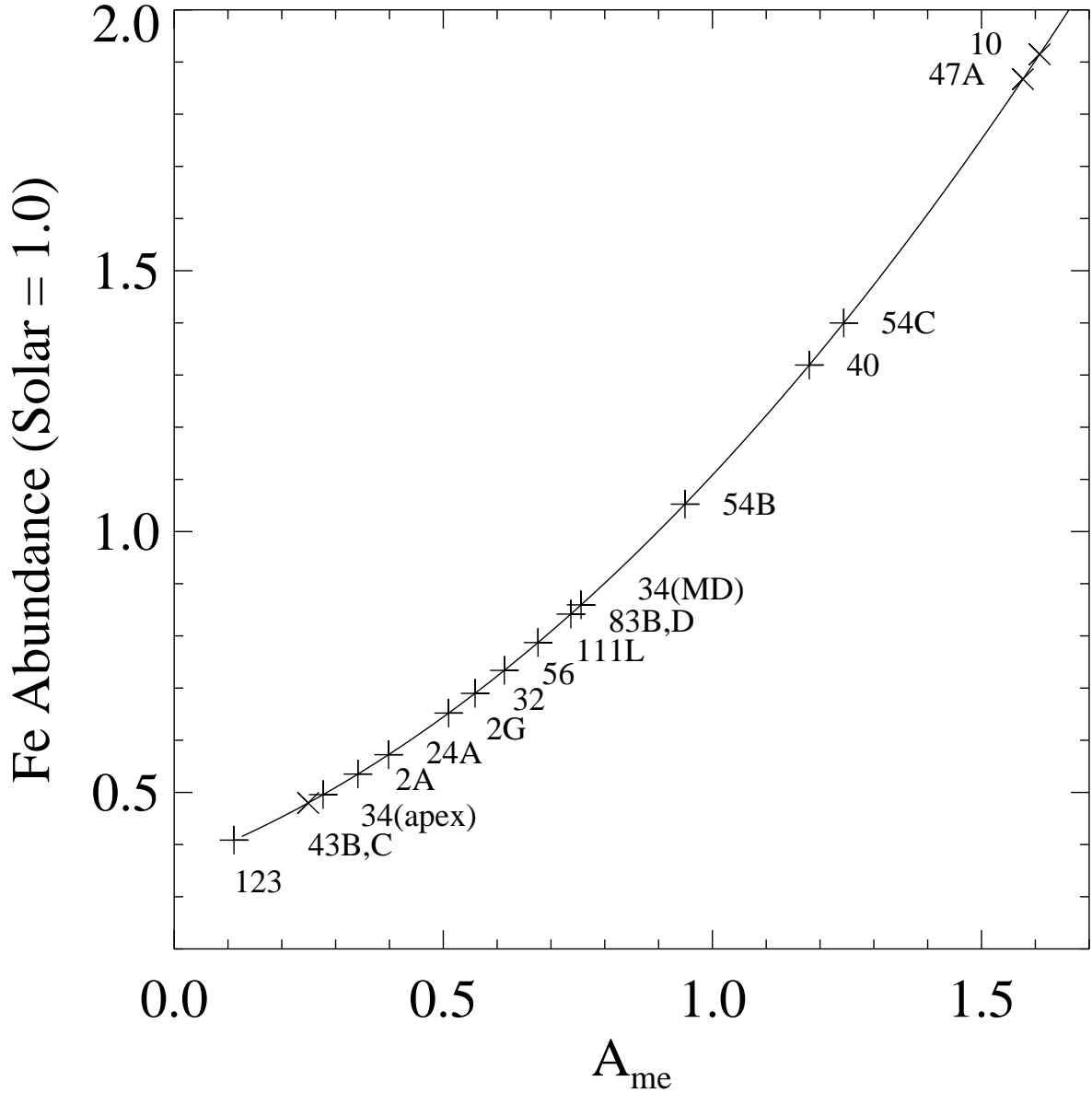}

\caption{The $A_{\rm me}$ values of 16 additional HH objects (derived
from data in the literature; see, e.g., the summary by Raga et al.\
1996) and the second order polynomial fit to the data in
Fig. \ref{fig1} (solid line) determine the approximate Fe abundance
for each object.  The symbols have the same meaning as in
Fig. \ref{fig1}, and the HH designation is indicated next to each
point. \label{fig2}}

\end{figure}

For the purpose of determining the approximate Fe abundance of
additional HH objects, we assume that all five objects studied by
BBN96 follow a one-dimensional relationship between $A_{\rm me}$ and
Fe abundance.  The data is well represented by a least squares fit to
a second order polynomial.  We assume, then, that the Fe abundance
(relative to the sun) of an HH object with an $A_{\rm me}$ value
(relative to HH 1) in the range of the abscissa of Figure \ref{fig1}
is approximately
\begin{equation} 
\label{fit} 
{\rm Fe \;\; abund.} \approx 0.36 + 0.39 A_{\rm me} + 0.36 A_{\rm me}^2
\end{equation}
This function overplots the data in Figure \ref{fig1} and is applied
to new $A_{\rm me}$ values in Figure \ref{fig2} (see \S
\ref{application}).  We want to point out that we have no theoretical
justification for the mathematical form of Equation \ref{fit}.
However, the qualitative results of this work (discussed below) are
not at all affected by the function chosen to fit the data.

\section{Fe Abundance Determination \label{application}}

We determined the $A_{\rm me}$ values for a number of additional HH
objects using the photometric data from the reference list put
together by Raga et al.\ (1996) in the footnote of their table 1.  The
additional objects comprise 13 high and 3 low excitation HH objects.
We would like to apply the correlation shown in Figure \ref{fig1} and
given in equation \ref{fit} to these objects.  To do so, we assume
that the earlier studied HH objects are somehow ``typical,'' or that
$A_{\rm me}$ is insensitive to the differences in the detailed
structure of each object.  Also, the new objects whose Fe abundances
we want to determine must have $A_{\rm me}$ values in the range of the
5 HH objects whose Fe abundances have been determined in detail.  The
objects studied in full detail cover the range from $A_{\rm me}$ =
0.143 (HH 43A) to 1.452 (HH 11), while the newly studied objects cover
the range from $A_{\rm me}$ = 0.111 (HH 123) to 1.601 (HH 10). In
other words, the newly added objects cover only a slightly larger
range of $A_{\rm me}$.  If we take into account that our selection of
objects was sort of arbitrary (based on the availability of spectra),
the fact that the range in $A_{\rm me}$ is relatively narrow may
already tell us that, even among our extended list of HH objects,
there is none whose gas phase Fe abundance deviates strongly from the
range of the 5 objects which were earlier studied in detail.
Consequently, we see no difficulty in this respect with applying our
approximate method.

\subsection{Gas Phase Fe Abundances for High Excitation HH Objects 
\label{heos}}

At first we restrict ourselves to the study of high excitation objects
because, in the present context, they are less problematic. As pointed
out above, high excitation objects are physically similar to each
other, so the correlation of $A_{\rm me}$ with the Fe abundance
(Fig. \ref{fig1}) is rather convincing.  We use Equation \ref{fit} and
the measured $A_{\rm me}$ values for our list of additional objects to
determine their Fe abundance.  The position of these objects in the
$A_{\rm me}$-Fe abundance diagram is shown in Figure \ref{fig2}.

\begin{figure}[t]
\plotone{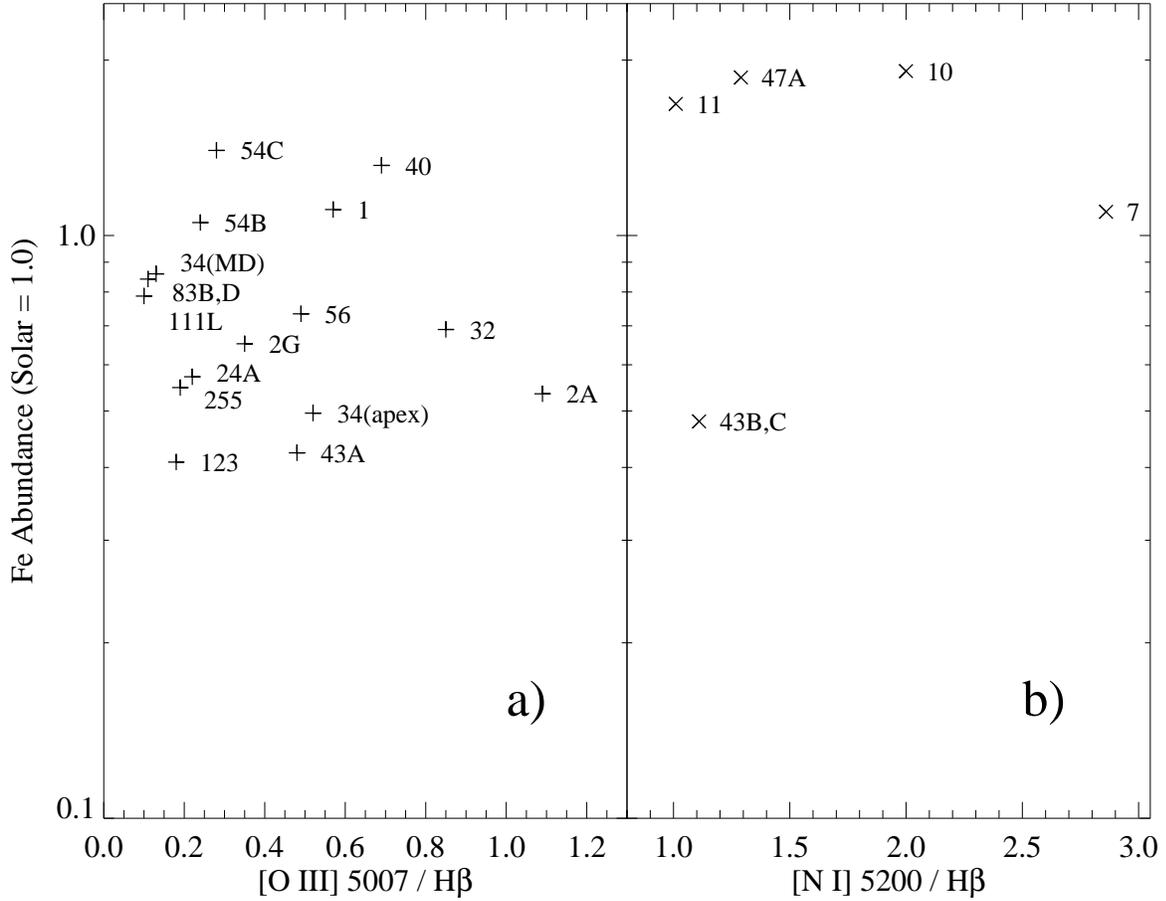}

\caption{Gas phase Fe abundance as a function of ``excitation'' (see
text, a measure of shock wave velocity) for high excitation (a) and
low excitation (b) HH outflows.  In (a), excitation increases to the
right, in (b) to the left.  No obvious correlation exists between the
Fe abundance of the 21 HH objects and their excitation.  The HH
designation is indicated to the right of each point. \label{fig3}}

\end{figure}

In Figure \ref{fig3} we have plotted the Fe abundance as a function of
the excitation of the HH object (i.e., as a function of the shock
velocity), which is most easily measured by the line flux ratio [O
III] 5007/H$\beta$ in high excitation objects (Fig. 3a) and by the
flux ratio [N II] 5200/H$\beta$ in low excitation objects (Fig. 3b).
The purpose of doing this is to determine whether the gas phase Fe
abundance depends on shock velocity.  If there occurs a drastic grain
destruction at some ``critical'' velocity (leading to a drastic
increase in the Fe abundance), there would be a discontinuity (or a
steep slope) in the Fe abundance in Figure 3.  A discontinuity could
also indicate that, at certain shock velocity, our approximate method
for the Fe abundance determination breaks down.

Figure \ref{fig3}, however, indicates no dependence of the Fe
abundance on the excitation.  The results for the high excitation
objects (Fig. 3a) show an abundance close to 1 but slightly higher
than 1 for 4 objects (HH 54B: 1.05, HH 1: 1.18, HH 40: 1.32, HH 54C:
1.42).  The other (12) objects show a (relatively modest) Fe gas phase
abundance depletion which seems to reach a limit of approximately a
factor 2.5 (HH 123 and HH 43A).  Our tentative conclusion is that, in
our sample, there are 4 objects which have basically unchanged
population I abundances for Fe (the reason being that dust grains have
been destroyed by shock waves and the Fe has gone back into the gas
phase). We leave it open, whether the very slight overabundance (5\% -
40\%) in these objects has to be considered as real or whether it may
be a consequence of our approximate method.  In addition to these
objects, there are also high excitation objects (including HH 123, HH
43A or HH 2A) where a certain fraction has gone back into the gas
phase, but a part may still be contained in dust. Thus, for high
excitation HH objects, it seems reasonable that there are no objects
for which a reasonably large fraction is still bound in dust.
However, if we take into account that the dust destruction is a very
sensitive function of the shock velocity \citep{mckeeea87}, it is
somewhat surprising that where there is depletion, it is never larger
than a factor of 2.5.
      
It is also very interesting to note that different parts of the same
HH objects (which may be due to different shock waves, like HH2A and
HH 2G; see Eisloeffel, Mundt \& B\"ohm 1994) tend to show the same gas
phase Fe abundances. This is obvious for HH 2A and HH 2G as well as
for HH 54B and HH 54C. Surprisingly, this is also true for the objects
HH 43A and HH 43B,C, where HH 43A is a high excitation and HH 43B,C
are low excitation objects (see Fig. 3a and 3b). It may however be
true that HH 43A and HH 43B,C form the apex and the ``wing'' of the
same bow shock (see the isophotic contour diagram of HH 43 presented
in B\"ohm \& Solf 1990, fig. 1), so that in this case, the result is
not too surprising.

Though there are some aspects of our Fe abundance results for high
excitation HH objects which we do not yet completely understand, we
can argue that the results in Figure 3a correspond approximately to
our expectations. As we shall now see the situation is not as
satisfactory when we consider the low excitation HH objects.

\subsection{Approximate Fe Abundance for Low Excitation HH Objects
\label{leos}}
 
We proceed in an analogous way to the one used for high excitation
objects. However, in this case we have available a total of only 5
objects, and the somewhat unexpected results have to be interpreted
with considerable caution. Of these 5 objects, 2 have been analyzed in
full detail for their gas phase Fe abundances (BBN96), namely HH 7 and
HH 11.  For the other 3 objects, we determined the Fe abundance using
the same procedure as for the high excitation objects (\S
\ref{heos}). The results are shown in Figures 2 and 3b.  In 3b, we
have plotted the results again as a function of excitation (shock wave
velocity), which is now (for the low excitation objects) measured by
the line ratio [N I] 5200/H$\beta$ (Raga et al.\ 1996) because the [O
III] line used in the high excitation objects is not emitted in the
low excitation objects. Our results show that (surprisingly) 4 out of
5 low excitation outflows show a gas phase Fe abundance which lies
somewhat above the solar Fe abundance (by a factor of 1.02 to 1.86).
There is only one low excitation HH object, namely HH 43B,C, which
shows some Fe depletion. But even in this case, the Fe depletion is
almost identical to the Fe depletion of the high excitation object HH
43A, with which it seems to be connected (see above).

The fact that 4 of the low excitation objects show an Fe abundance
somewhat larger than 1 may not be significant because our approximate
method may not be very accurate, especially for low excitation
objects.  It is possible that the Fe abundance in these objects
corresponds reasonably well to the solar abundance.  We are, however,
surprised that 4 out of 5 low excitation HH objects show no depletion
of the gas phase Fe abundances. We would have (naively) expected that
these objects show strong Fe depletion (by a factor of roughly 100)
because no dust grains would have been destroyed in the weak shock
waves of the low excitation objects. We would expect to see the
original gas phase composition of the molecular cloud. We must
conclude that the matter which is visible now in a low excitation HH
object went through a shock wave of a much higher shock velocity at an
earlier time.

\section{Conclusions \label{conclusions}}

Earlier detailed studies of the gas phase Fe abundance for a very few
HH objects had led to the conclusion that there is no simple
correlation between the Fe abundance and the HH shock velocity (BBN94,
BBN96).  A correlation would have been expected if the molecular cloud
matter enters the HH working surface and dust grains are destroyed in
the HH bow shock by the usual processes (mostly by non-thermal
sputtering, see e.g., McKee et al.\ 1987; Draine 1995). Since the
study of BBN96 is based on only 5 objects (HH 1, HH 7, HH 11, HH 43A,
HH 255), and adding more objects to this list using a detailed method
requires great effort, we have derived Fe abundances of 16 additional
HH objects using a very approximate method.
     
We introduced a characteristic number, $A_{\rm me}$ (Eqn. \ref{fit}),
which depends on the (frequently measured) line fluxes of [Fe II] 5159
and [Fe II] 7155, as well as on the [Ca II] 7291 line flux and the
H$\alpha$ and H$\beta$ fluxes.  For the five HH objects (mentioned
above) for which we have a detailed gas phase analysis, $A_{\rm me}$
is correlated reasonably well with the gas phase Fe abundance. If we
assume that this correlation is not only approximately correct for the
5 objects studied in detail, but for all HH objects for which we can
determine $A_{\rm me}$, then we can use this correlation to determine
their gas phase Fe abundance.

For the high excitation HH objects, our results are the following.  In
addition to the 3 high excitation objects for which the Fe gas phase
abundances have been measured in detail (BBN96), we have now
determined the Fe gas phase abundance for 13 more objects. Of these 16
objects, 4 show a gas phase Fe abundance between 1.0 and 1.44 times
the solar abundance. Taking into account the possible errors of the
abundance determination, this probably means that 4 of these objects
have unchanged population I Fe gas phase abundances (no Fe bound in
dust grains). The other 12 high excitation objects show some (but
relatively small) Fe depletion which never gets larger than a factor
of 2.5.

In general, the result looks reasonable. The high excitation HH
objects show that, in some cases, Fe has gone back completely to the
gas phase, and in other cases, some (but not too much) Fe may still be
bound in dust grains, leading to moderate depletion of the gas phase
abundance. We also find that similar depletion exists in different
parts of the same object (e.g., in HH 2A and HH 2G or in HH 54B and HH
54C).  The gas phase Fe abundance for high excitation objects agrees
qualitatively with the predictions given, e.g., by McKee et al.\
(1987) that dust grains are destroyed by strong shocks. The limit of
the depletion factor of 2.5 is interesting but has not been explained
in detail.

Understanding the results for the low excitation HH objects seems to
be more difficult, and some questions remain open. The first problem
is that we could study only 5 objects. For these objects we find quite
unexpected results. Four of them have an Fe gas phase abundance
between 1.02 and 1.86 times the solar Fe abundances.  One may argue
that our uncertainty (which is supposedly larger in the low excitation
than in the high excitation objects) might permit these to be
consistent with 1 solar Fe abundance.  However, we still have the
problem that low excitation objects (for which we do not expect
destruction of dust grains) apparently have all or most of their Fe in
the gas phase.

It is also important that we find no correlation of the Fe abundance
with shock velocity.  While this is not surprising for the high
excitation objects, we would have (naively) expected a large
difference between the low and high excitation objects.  Specifically,
we would have expected a strong Fe depletion (perhaps by a factor of
100) in the low excitation objects, since their shocks are not strong
enough to destroy grains \citep{mckeeea87,draine95}.  The (perhaps)
simplest explanation is that the material we have studied in low
excitation objects has passed through a faster shock at earlier times.
This may be a surprising result, but we cannot see any way to avoid
it.

It is difficult to speculate about this process.  What would be
required is that the matter that enters the low excitation bow shock
(which we assume to be molecular cloud matter) has previously gone
through a high excitation bow shock (in which the dust grains were
destroyed).  The simplest assumption would be that the previous high
excitation bow shock was generated by the same source star that
generates the low excitation HH outflow.  This suggests that none of
the five low excitation objects studied in this work correspond to the
leading edge of the stellar outflow (i.e., all of them are produced
some time after the onset of the jet).

It is also interesting to note that we have assumed the molecular
clouds in which these HH objects reside have approximately solar Fe
abundance (in gas and dust phases).  Though this is a reasonable
assumption, it may be possible that the abundances in present-day
molecular clouds could differ from solar by a factor of a few.  The
entire range of Fe abundances of the objects we have studied is from
$\sim 0.4$ to 1.9 times solar.  In light of this, one may argue that
the dust grains are completely destroyed (by passing through fast
shocks) in all 21 objects, and that the observed spread in Fe
abundance is due to different abundances in their parent molecular
clouds.  At present, our data cannot resolve this issue.

\acknowledgments

We thank the referee, Bo Reipurth, for carefully reading the
manuscript and making very useful suggestions for improvement.  This
research has been supported by NSF grant AST-9729096.


\end{document}